\documentclass{iopart}
\usepackage{graphicx}

\begin{document}

\title{Avalanche Dynamics for Active Matter in Heterogeneous Media}
 
\author{C. J. O. Reichhardt and C. Reichhardt}
\address{Theoretical Division and Center for Nonlinear Studies,
Los Alamos National Laboratory, Los Alamos, New Mexico 87545, USA}
\ead{cjrx@lanl.gov}

\begin{abstract}
Using numerical simulations, we examine the dynamics of active matter run-and-tumble disks moving in a disordered array of obstacles. As a function of increasing active disk density and activity, we find a transition from a completely clogged state to a continuous flowing phase, and in the large activity limit, we observe an intermittent state where the motion occurs in avalanches that are power law distributed in size with an exponent of $\beta = 1.46$.  In contrast, in the thermal or low activity limit we find bursts of motion that are not broadly distributed in size.  We argue that in the highly active regime, the system reaches a self-jamming state due to the activity-induced self-clustering, and that the intermittent dynamics is similar to that found in the yielding of amorphous solids.  Our results show that activity is another route by which particulate systems can be tuned to a nonequilibrium critical state.  
\end{abstract}

\maketitle

\vskip 2pc

\section{Introduction}
There are numerous examples of driven 
collectively interacting systems that exhibit avalanches or intermittent behavior
when driven over quenched 
disorder, including  vortices in type-II superconductors
\cite{1,2,3}, magnetic domain walls \cite{4,5}, 
earthquake models \cite{6,7}, and colloidal depinning over rough landscapes \cite{8}.   
At a critical driving force $F_{c}$, there is
a depinning transition from a pinned to a sliding state.
Motion often occurs in avalanches
close to the depinning transition,
and if depinning is associated with critical features such as diverging
characteristic lengths and times,
the avalanches and other fluctuating quantities will exhibit broad or power law
distributions \cite{5,7}.
Scale-free avalanche dynamics often appear
near yielding or unjamming  
transitions,
such as in the
intermittent motion of 
dislocations in crystalline solids \cite{9,10,11} 
or the rearrangements of particles at yielding in amorphous 
materials \cite{12,13,14,15}. 
For loose assemblies of particles such as grains or bubbles,
the shear modulus becomes finite above a density $\phi_j$ when a
jamming transition occurs \cite{16,17},
and it is known that in such systems, the dynamics become
increasingly intermittent
as the jamming point is approached, 
producing power law distributions in
a variety of dynamic quantities \cite{18,19,20,21,22,23}.   

In the driven systems described above, the dynamics arise from
some form of externally applied driving or shear.
In contrast, in active matter systems the particles are {\it self}-driven.
Examples of active matter systems that have been attracting increasing attention
include
pedestrian flow, biological systems such 
as run-and-tumble bacteria, and self-propelled colloids \cite{24,25}.
The behavior of these systems can
be captured by a simple model consisting of sterically interacting hard disks
with a self-mobility represented either by
driven diffusion
or run-and-tumble dynamics.
In two dimensions, a non-active, thermal assembly of hard disks
forms a uniform density liquid at finite temperature, and if the
disk density is large enough, a jammed or crystalline state emerges \cite{17}.
If the disks are self-propelled or active,
for large enough activity
a transition can occur 
from
the uniform liquid state to a phase 
separated state in which a high density cluster that can be regarded as a solid is
surrounded by a low density gas of disks,
even when the overall density of the system
is well below that at which non-active disks would jam or crystallize \cite{26,27,28}.
Self-clustering occurs when multiple active disks 
collide and continue to swim into each other, producing an active load-bearing
contact in a system containing no tensile forces, and it has
been observed in experiments
using self-propelled colloids \cite{29,30}
and in simulations of
disks obeying driven diffusive or 
run-and-tumble dynamics \cite{28}.
Previous numerical studies \cite{31} of
active run-and-tumble disks driven though random obstacle arrays 
show that for fixed active disk density,
the average drift mobility of the disks is a nonmonotonic function of the
activity,  initially increasing with increasing run length, but then passing
through a maximum and decreasing at large run lengths,
with the onset of self-clustering or self-jamming coinciding with the mobility reduction.
For a fixed run length, the mobility decreases
as the active disk density increases
due to crowding effects.  

In this work we examine the motion of active run-and-tumble disks
driven through a random obstacle
array as a function of obstacle density and activity. 
We find that
there is a critical amount of disorder and activity
above which
the motion becomes highly intermittent and takes the form of avalanches that
are power law distributed in size,
$P(s) \propto s^{-\beta}$, with an exponent of $\beta = 1.46 \pm 0.1$.
For fixed 
obstacle density but decreasing activity,
the motion becomes more continuous, the avalanche behavior is lost, and
the disks act like a fluid moving through the obstacle array,
while at zero or very small activity,
the disks become completely clogged and there is no motion.  
We argue that
in the limit of large activity, critical behavior occurs
due to self-jamming or self-clustering, while
in the low but finite activity limit the
disks act like a liquid with continuous fluctuations.
The critical behavior
under an external drive can be viewed
as analogous to a yielding transition of an amorphous solid close to
a jamming point.
We also find that for fixed active disk density,
a critical amount of disorder in the form of obstacles must be added to produce
power law distributed avalanche sizes,
similar to the behavior observed for
avalanches in certain magnetic systems
\cite{5,32,33}.
Our results indicate that activity can provide another method for tuning
a system to a nonequilibrium critical state.

\begin{figure}
  \center
\includegraphics[width=0.8\columnwidth]{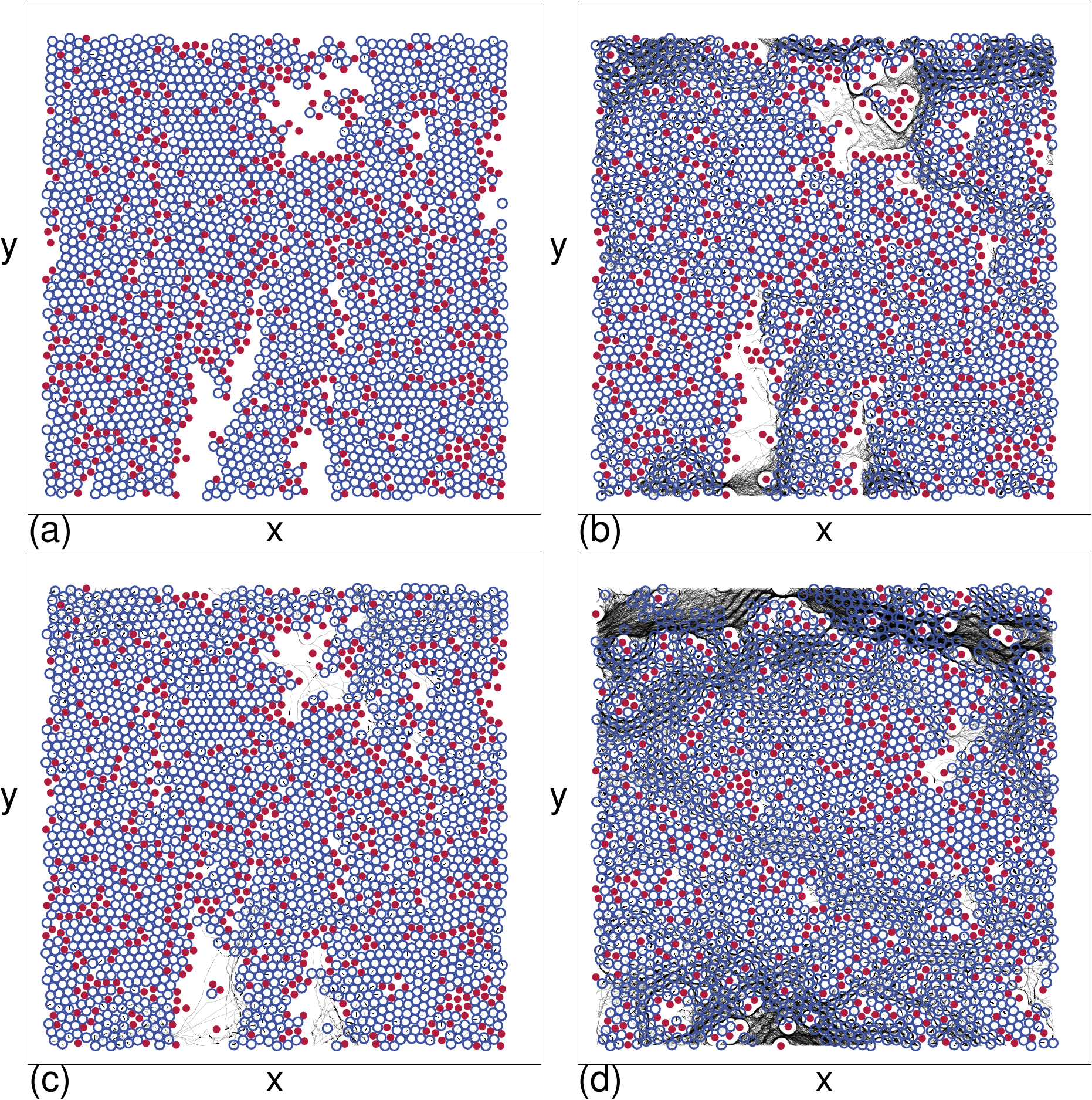}
\caption{ The obstacle positions (red filled circles),
  active disks (dark blue open circles), and trajectories (light blue lines)
  for a system with $\phi_{\rm tot} = 0.754$.
  (a) For low activity $l_{r} = 0.1$ at $\phi_{\rm obs}=0.1727$,
   the disks become completely
  clogged and there is no motion.
  (b) For intermediate activity $l_{r} = 1$ at $\phi_{\rm obs}=0.1727$,
  we find a combination of
  flowing and immobile active disks.
(c) For high activity $l_{r} = 320$ at $\phi_{\rm obs}=0.1727$, 
  motion occurs only in bursts or avalanches.
  (d) A system with $l_{r} = 320$ at a lower 
obstacle density of $\phi_{\rm obs} = 0.1256$.
}
\label{fig:1}
\end{figure}

\section{Simulation}
We simulate a two-dimensional $L \times L$ system with
$L=50$ and with periodic boundary conditions in the $x$ and $y$-directions
containing
$N_{\rm obs}$ obstacles and
$N_{a}$ active disks of radius $R=0.5$. 
The obstacles are identical to the active disks except their locations are
permanently fixed.
Steric disk-disk interactions are given by a harmonic repulsive force
${\bf F}_{dd} = k(d-2R)\Theta(d - 2R){\hat {\bf d}}$ 
where $d$ is the distance and ${\hat {\bf d}}$ is the displacement vector between
a pair of disks.
The spring constant $k = 100$
is large enough that, for the parameters we consider,
disks overlap by less than $0.01R$,
so the system is approximately in the hard disk limit.
The obstacle area coverage is $\phi_{\rm obs} = N_{\rm obs}\pi R^2/L^2$,
the active disk area coverage is $\phi_{a} = N_{a}\pi R^2/L^2$, and the total 
area coverage is $\phi_{\rm tot} = \phi_{\rm obs} + \phi_{a}$. 
In an obstacle-free system with $\phi_{\rm obs} = 0$,
in the absence of activity
the disks form a hexagonal solid near $\phi_{a} = 0.9$.
The active disk dynamics obey the following overdamped
equation of motion: 
\begin{equation}
\eta \frac{d{\bf r}_i}{dt} = {\bf F}^i_{\rm inter} + 
{\bf F}^i_{m} + {\bf F}^i_{\rm obs} + {\bf F}_{D}
\end{equation}
where the damping coefficient $\eta = 1$.
The interactions between active disks
are given by ${\bf F}^i_{\rm inter}=\sum_j^{N_a}{\bf F}_{dd}^{ij}$, and  
${\bf F}^i_{m}=F_m{\bf \hat m}$ is a
run-and-tumble motor force with $F_m=0.5$ that acts in a randomly
chosen running direction ${\bf \hat m}$ for a running time $\tau$, after which a new
running direction ${\bf \hat m}^\prime$ is randomly chosen.
In the absence of any collisions, during the running time an active disk
moves a run length
$l_{r} = F_{m}\tau \delta t$, where $\delta t=0.002$ is the simulation time step.
The obstacle forces are given by ${\bf F}_{\rm obs}=\sum_{k}^{N_{\rm obs}}F_{dd}^{ik}$, 
and the external driving force ${\bf F}_{D}=F_D{\bf \hat x}$ is  
applied uniformly to all active disks with $F_D=0.5$.
To initialize the system, we place a density $\phi_{\rm tot}$ of disks at
nonoverlapping locations in the sample, and then randomly
choose $N_{\rm obs}$ of the disks to serve as obstacles, fixing them in their
original random locations.
We apply a driving force ${\bf F}_{D}$   
and wait several million simulation time steps to ensure that we have reached
a steady state before
measuring the active disk velocity fluctuations and displacements.
We obtain a time series of the average active
disk velocity 
in the driving direction,
$V(t) = N_{a}^{-1}\sum^{N_a}_{i} v_{i}^x(t)$,
where $v_i^x(t)={\bf v}_i(t)\cdot {\bf \hat x}$,
and also measure the time-averaged active disk velocity
in the driving direction $\langle V \rangle=\langle V(t)\rangle$.
We quantify the activity level using
$l_{r}$ and the disorder using $\phi_{\rm obs}$.    
      
\section{Results}

\begin{figure}
  \center
\includegraphics[width=0.6\columnwidth]{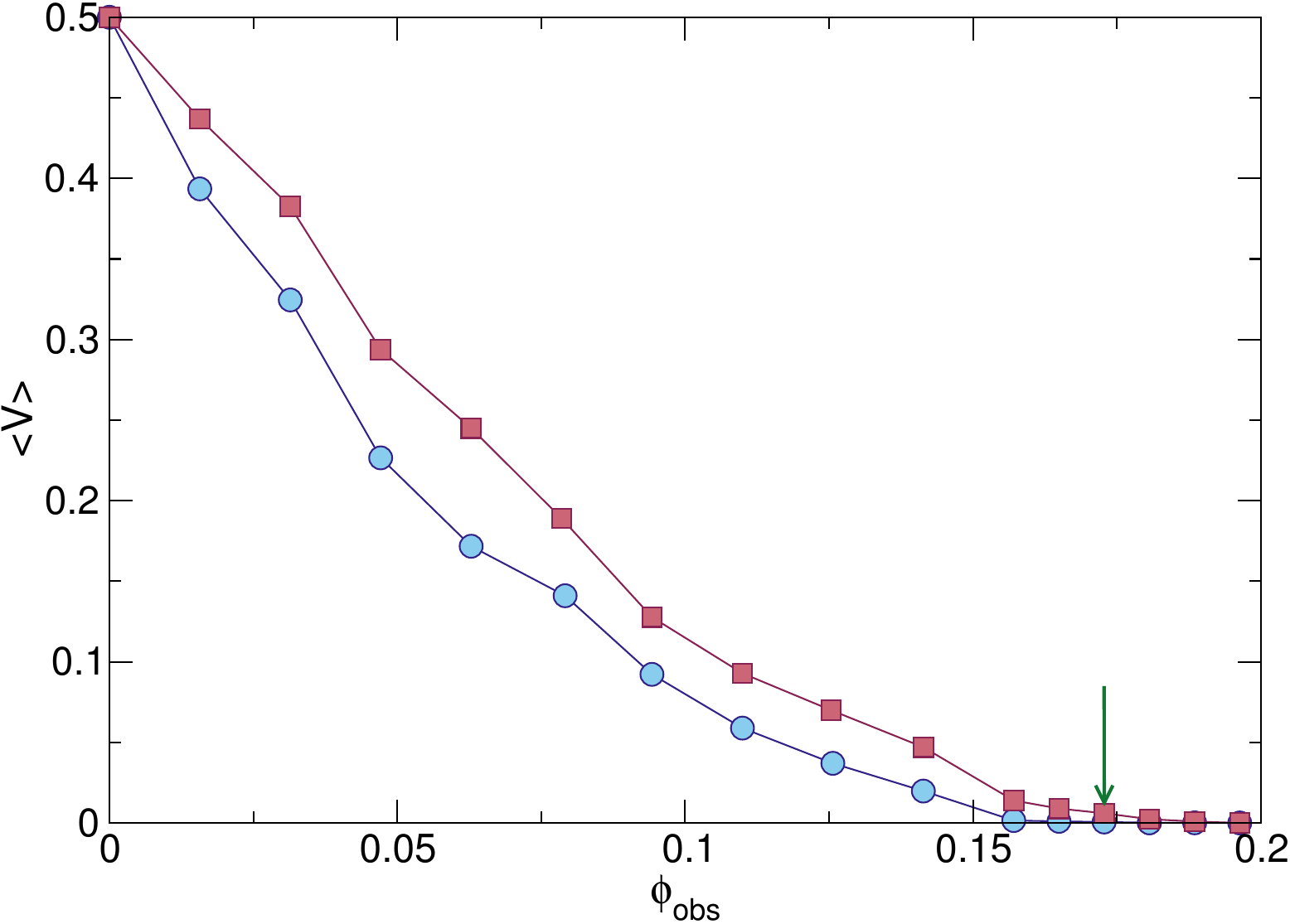}
\caption{ Average active disk velocity $\langle V\rangle$ in the direction of
  the applied drive vs
  $\phi_{\rm obs}$ for a system with
  $\phi_{\rm tot} = 0.754$ at
  $l_{r} = 1$ (red squares) and $l_{r} = 320$ (blue circles).
  The green arrow indicates the value of $\phi_{\rm obs}$ used in Fig.~\ref{fig:3}.
}
\label{fig:2}
\end{figure}

In Fig.~\ref{fig:1} we illustrate the behavior of a system with
$\phi_{\rm tot} = 0.754$.
For $\phi_{\rm obs} = 0.1727$,
Fig.~\ref{fig:1}(a) shows that at a very low activity level of
$l_r=0.1$, 
the disks become completely clogged 
with $\langle V\rangle = 0$.
In Fig.~\ref{fig:1}(b)
at a higher level of activity $l_{r} = 1$,
we find
a coexistence of jammed and moving active disks,
while for large activity $l_r=320$ in Fig.~\ref{fig:1}(c),
the flow is highly intermittent and occurs through 
avalanches.
Figure~\ref{fig:1}(d)
shows that in an $l_r=320$ system with 
a lower obstacle density of
$\phi_{\rm obs} = 0.1256$, the flow of active disks becomes continuous again.

\begin{figure}
  \center
\includegraphics[width=0.8\columnwidth]{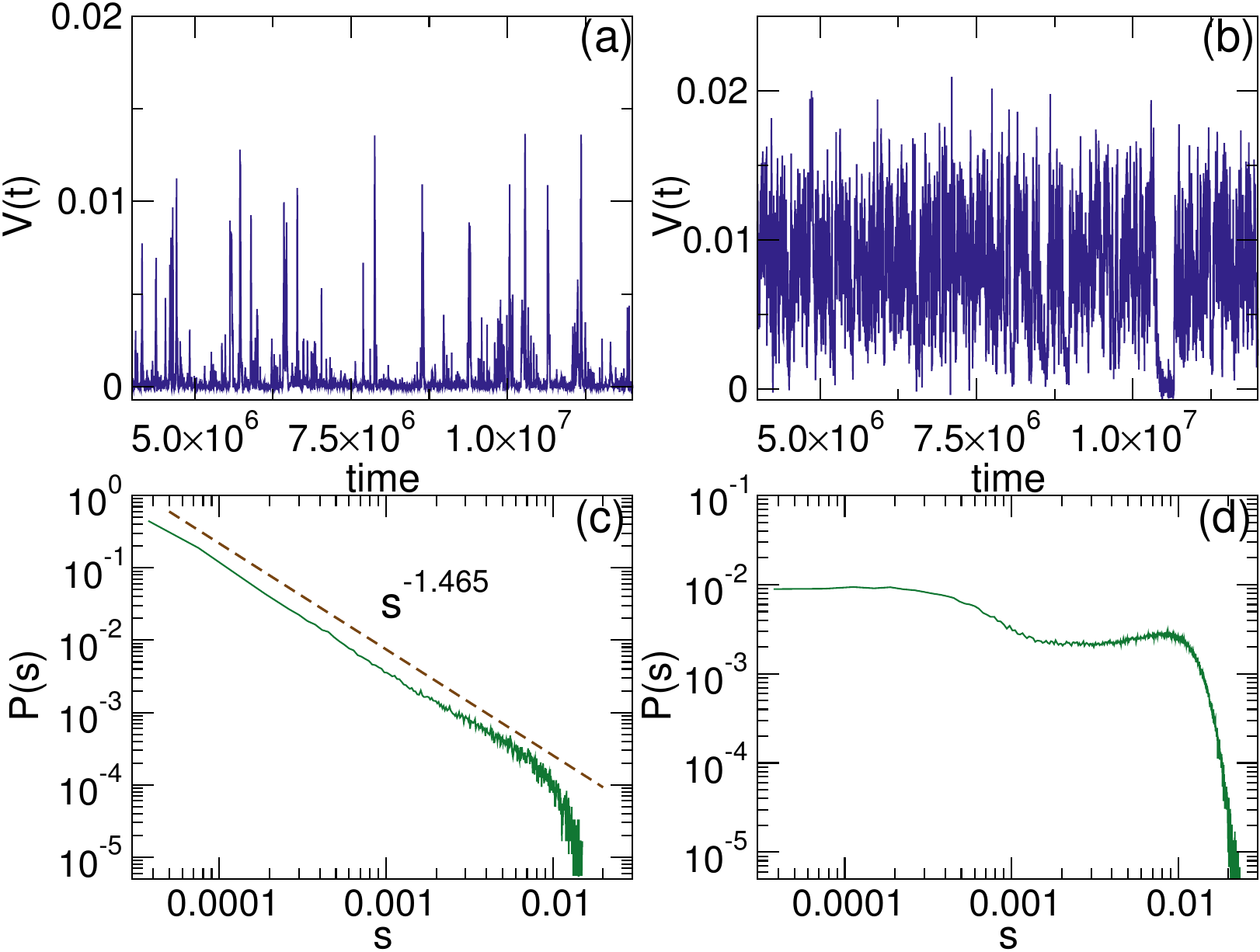}
\caption{ (a) A portion of the time series of the average
  disk velocity $V(t)$ from the system in 
  Fig.~\ref{fig:2} at $\phi_{\rm tot}= 0.754$,
  $\phi_{\rm obs} = 0.1727$, and $l_{r} = 320$, where the
  motion is strongly intermittent.
  (b) The avalanche size distribution $P(s)$ from the complete time series for the
  system in panel (a).
  The dashed line is a power law fit with exponent
  $\beta = 1.465 \pm 0.15$.
  (c) $V(t)$ for the $l_r=1$ system from Fig.~\ref{fig:2}.
  (d) $P(s)$ for the system in panel (c) has a bimodal rather than a power law shape.
}
\label{fig:3}
\end{figure}

In Fig.~\ref{fig:2} we plot the average drift velocity
per active disk in the direction of drive $\langle V\rangle$ versus $\phi_{\rm obs}$
over the range $0 \leq \phi_{\rm obs} \leq 0.196$
for a system with $\phi_{\rm tot} = 0.754$ at
$l_{r} = 1$ and $l_r=320$.
As shown in previous work \cite{34},
self clustering occurs when $l_{r} > 10$
for $\phi_{\rm tot}=0.754$,
so the values of $l_{r}$ in Fig.~\ref{fig:2}
are representative of  
the liquid state and the phase separated state.    
At $\phi_{\rm obs}= 0$ the disks undergo free flow drift giving 
$\langle V\rangle = F_{D} = 0.5$ for all $l_r$;
however, as $\phi_{\rm obs}$ increases, 
$\langle V\rangle$ is always lower
in the phase separated $l_r=320$ sample than in the
liquid $l_r=1$ sample.
For $\phi_{\rm obs} > 0.15$, the motion in
the $l_{r} = 320$ system becomes highly intermittent, as shown
by the plot of $V(t)$ 
in Fig.~\ref{fig:3}(a)
for the system in Fig.~\ref{fig:2} with $l_r=320$ at
$\phi_{\rm obs}=0.1727$, which is also illustrated in Fig.~\ref{fig:1}(c).
Here the motion occurs in bursts or avalanches.  
In contrast, the $l_r=1$ sample at the same obstacle density has a much more
continuous $V(t)$, as shown in Fig.~\ref{fig:3}(b) and illustrated in Fig.~\ref{fig:1}(b).
The average velocity ratio $\langle V\rangle_{l_r=1}/\langle V\rangle_{l_r=360}=17$
for $\phi_{\rm obs}=0.1727$, indicating how strongly an increase in activity can
reduce the flow through the system.
We use the time series $V(t)$ to construct an avalanche size distribution $P(s)$, where
$s$ is defined to be equal to the instantaneous value of $V$.
In Fig.~\ref{fig:3}(b), $P(s)$ for the $l_r=320$ system can be fit to a power law
distribution over two decades
with an exponent $\beta=1.465\pm 0.15$, while in Fig.~\ref{fig:3}(d), $P(s)$ for
the $l_r=1$ system is not broad but has a bimodal distribution, where the
second peak is characteristic of the flow of a liquid through a disordered
medium \cite{8}.

\begin{figure}
  \center
\includegraphics[width=0.8\columnwidth]{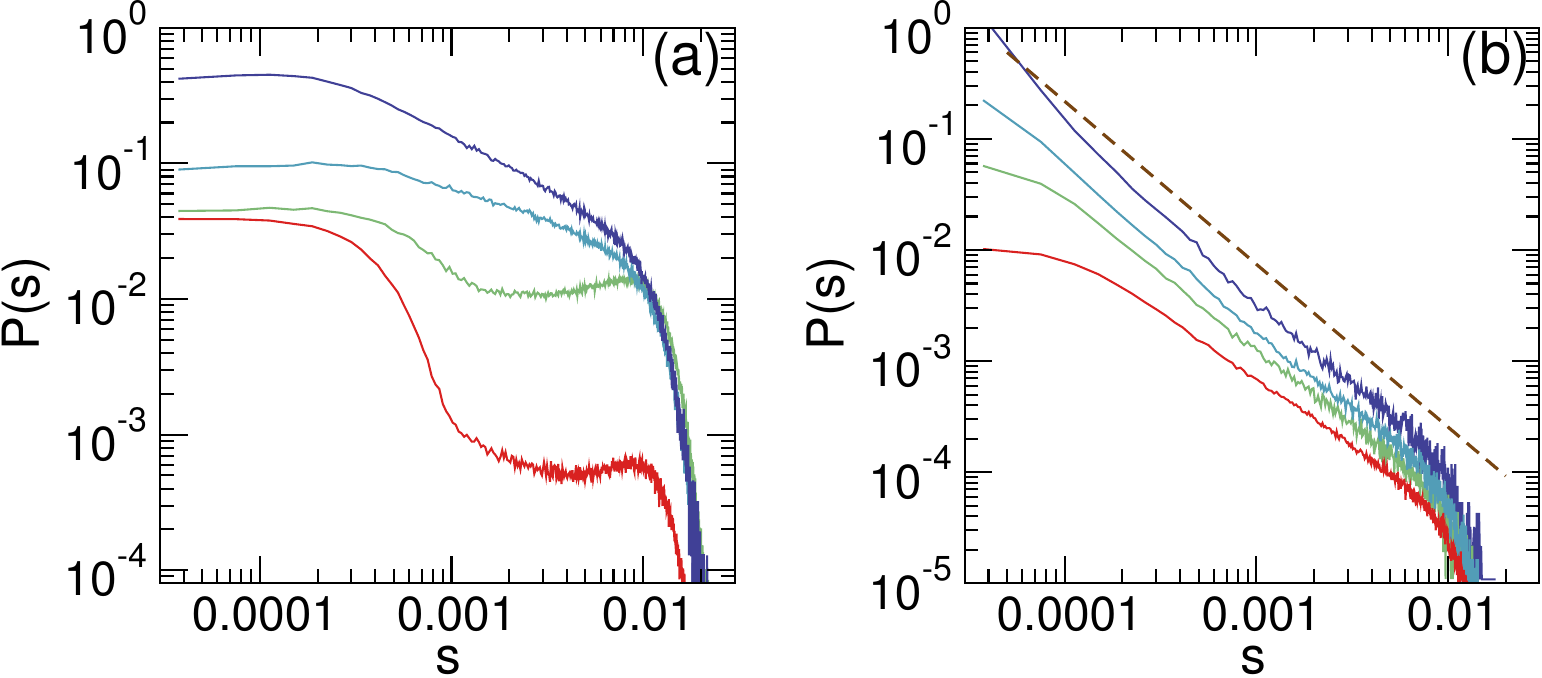}
\caption{ Avalanche size distribution $P(s)$
  for a system with $\phi_{\rm tot} = 0.754$ and $\phi_{\rm obs} = 0.1727$.
  (a) $l_{r} = 0.3$ (red),
  1.0 (green), 10 (light blue), and 20 (dark blue), from bottom to top.
  The curves have been shifted vertically for clarity.
  (b) $l_{r} = 80$ (red), 160 (green),  320 (light blue),  and $640$ (dark blue),
  from bottom to top.  The curves have been shifted vertically for clarity.
  The dashed line indicates a power law fit with exponent
  $\beta = 1.465$.
}
\label{fig:4}
\end{figure}

In Fig.~\ref{fig:4}(a) we plot
$P(s)$ for a system with  $\phi_{\rm tot} = 0.754$ and $\phi_{\rm obs} = 0.1727$ 
at $l_{r} = 0.3$, 1.0, 10, and 20.
The bimodal characteristics of the $l_{r} = 0.3$ and $l_r=1.0$ distributions are
lost for $l_r=10$ when the system acts like a fluid.
The avalanche size distribution broadens when self-induced clustering
begins to occur,
and for $l_{r} = 20$ and above it is possible to fit a power law to a portion of $P(s)$.
Fig.~\ref{fig:4}(b) shows $P(s)$ for the same system at
$l_{r} = 80$, 160, 320, and $640$ along with a dashed line indicating a power law
fit with exponent
$\beta = 1.465$.
The region over which $P(s)$ obeys a power law grows in extent as
$l_r$ increases, and the exponent falls in the range $1.35 \leq \beta \leq 1.5$.
The overall shape of $P(s)$ remains nearly the same for $l_r=80$ and above,
but the time intervals separating successive avalanche events increase with
increasing $l_r$.
These
results indicate that there is a critical $r_{l}$ above which
scale-free avalanches occur, and that this critical value corresponds to the
point at which the system begins to act like a solid rather than a liquid.

\begin{figure}
  \center
\includegraphics[width=0.8\columnwidth]{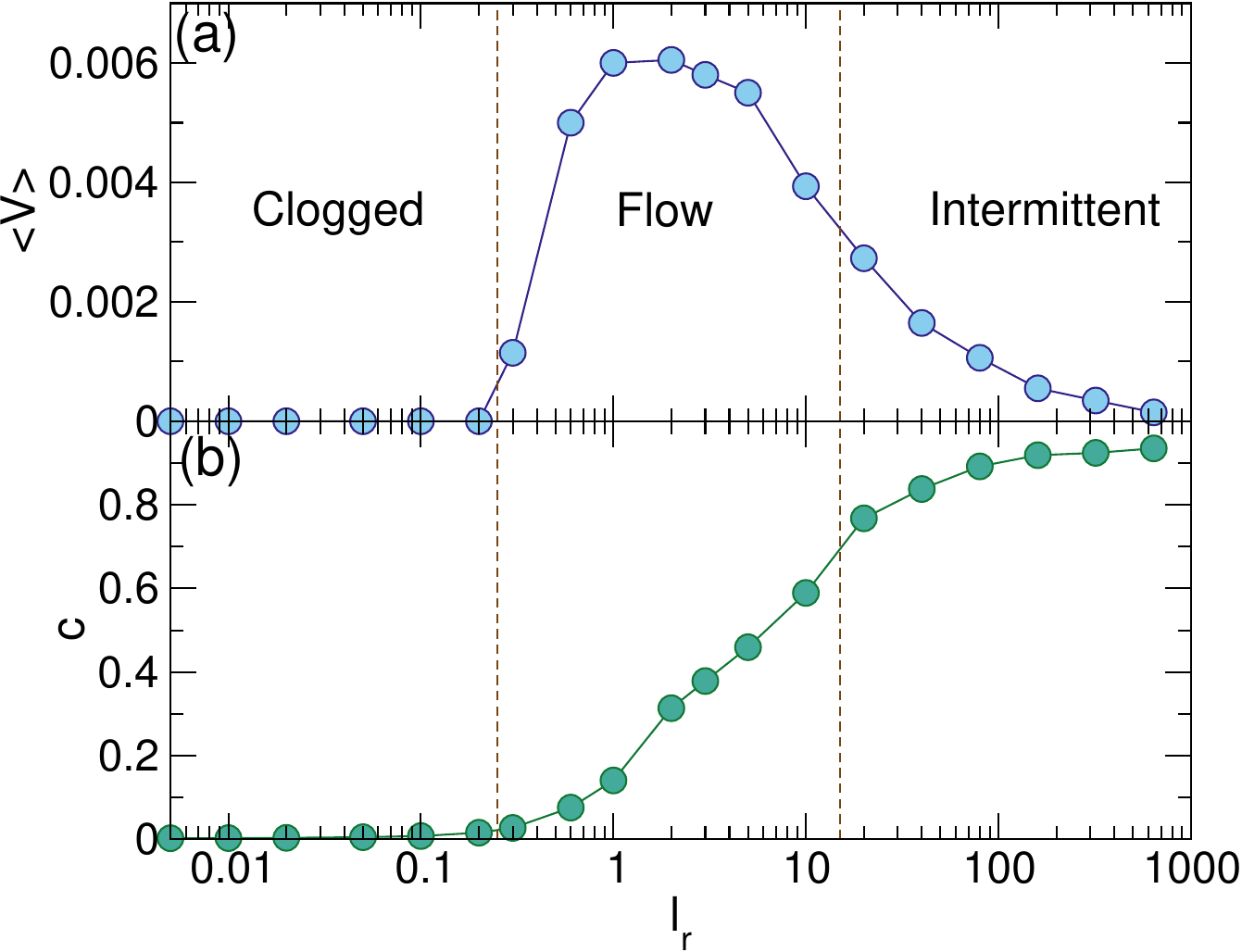}
\caption{The time-average disk flux through the system
  $\langle V\rangle$ vs $l_{r}$ for
  samples with $\phi_{\rm tot} = 0.754$ and $\phi_{\rm obs} = 0.1727$
  showing the completely clogged regime at low $l_r$,
  the flow regime at intermediate $l_r$, and the intermittent or avalanche regime
  at large $l_r$. 
  (b) The fraction $c$ of disks that are in the largest cluster
  vs $l_{r}$ for a system with $\phi_{\rm tot}=0.754$ and $\phi_{\rm obs}=0$,
  showing that the onset of the intermittent phase
  correlates with the onset of self-clustering.
}
\label{fig:5}
\end{figure}
 
In Fig.~\ref{fig:5}(a) we plot $\langle V\rangle$ versus $l_{r}$
for the system in Fig.~\ref{fig:2} at $\phi_{\rm tot} = 0.754$
and $\phi_{\rm obs}= 0.1727$.
Based on the behavior of the $P(s)$ distributions, as illustrated in Fig.~\ref{fig:4},
we identify three regimes:
a fully clogged state for $l_{r} \leq 0.2$,
where $\langle V\rangle = 0$ as illustrated in Fig.~\ref{fig:1}(a),
a flowing or liquidlike region for
$ 0.2 < l_{r} < 20$,
and an intermittent avalanche regime for $l_{r} \geq 20$.
We measure the fraction $c$ of disks that belong to the largest cluster using
the cluster identification algorithm described in \cite{hermann}, and
in Fig.~\ref{fig:5}(b) we plot $c$  versus
$l_{r}$ for an obstacle-free sample with the same $\phi_{\rm tot}=0.754$ as in
Fig.~\ref{fig:5}(a) but with $\phi_{\rm obs}=0$,
showing that the onset of the intermittent phase in the presence of obstacles
correlates with a large increase in self-clustering  in the absence of obstacles.

\begin{figure}
  \center
\includegraphics[width=0.8\columnwidth]{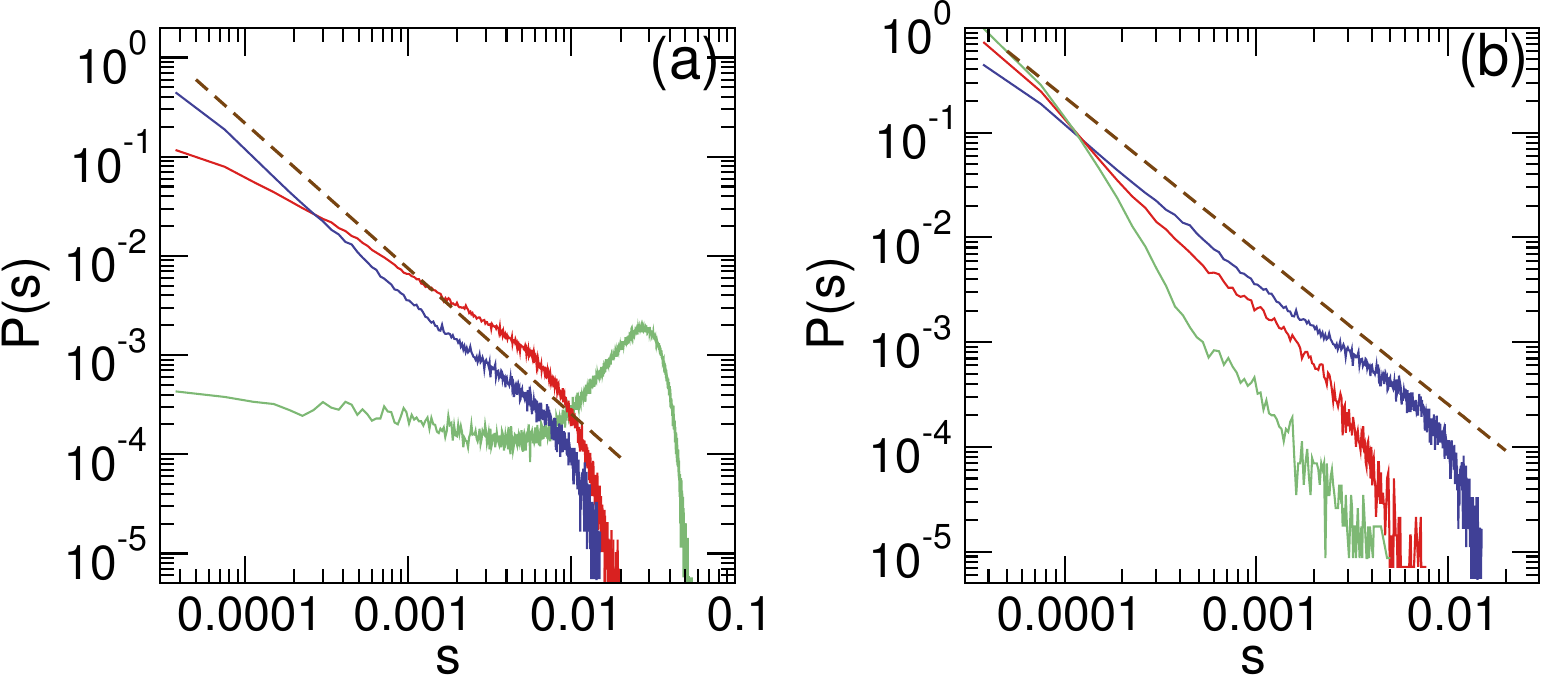}
\caption{
  Avalanche size distributions $P(s)$
  for a system with $\phi_{\rm tot} = 0.754$ and $l_{r} = 320$ at
  (a) $\phi_{\rm obs} = 0.1413$ (green), $0.157$ (red) and $0.1727$ (blue).
  (b) The same for $\phi_{\rm obs} = 0.1727$ (blue),
$0.1884$ (red), and $0.2$ (green).
  The dashed lines are power law fits with
  $\beta = 1.465$.
}
\label{fig:6}
\end{figure}

The avalanche size distributions are also affected by the obstacle density.
In Fig.~\ref{fig:6}(a) we plot $P(s)$
for fixed $\phi_{\rm tot} = 0.754$ and $l_{r} = 320$ 
at $\phi_{\rm obs} = 0.1413$, $0.157$, and $0.1727$.
For $\phi_{\rm obs} = 0.1413$, even though the system can self-cluster there is enough  
room for the disk clusters to flow freely around the obstacles,
giving a peak at a characteristic avalanche size $s \approx 0.03$,
while at $\phi_{\rm obs}  = 0.157$, avalanches of size $s>0.2$ are lost and
$P(s)$
begins to broaden.
In Fig.~\ref{fig:6}(b)
we plot $P(s)$ for the same system with
$\phi_{\rm obs} = 0.1727$, $0.1884$,  and $0.2$.
The maximum avalanche size $s_{\rm max}$ continues to decrease
as the obstacle density increases,
and at $\phi_{\rm obs} = 0.2041$ the avalanche motion
is completely suppressed
since $\langle V\rangle = 0$.
This behavior is similar to what has 
been predicted for
models of avalanches in magnetic systems,
where it is necessary to add a critical amount of disorder in order
to obtain avalanches that are power law distributed in size
\cite{5,32,33}.
For weak disorder the
magnetic avalanches are dominated
by system spanning events, while for strong disorder only
small avalanches occur.

\begin{figure}
  \center
\includegraphics[width=0.6\columnwidth]{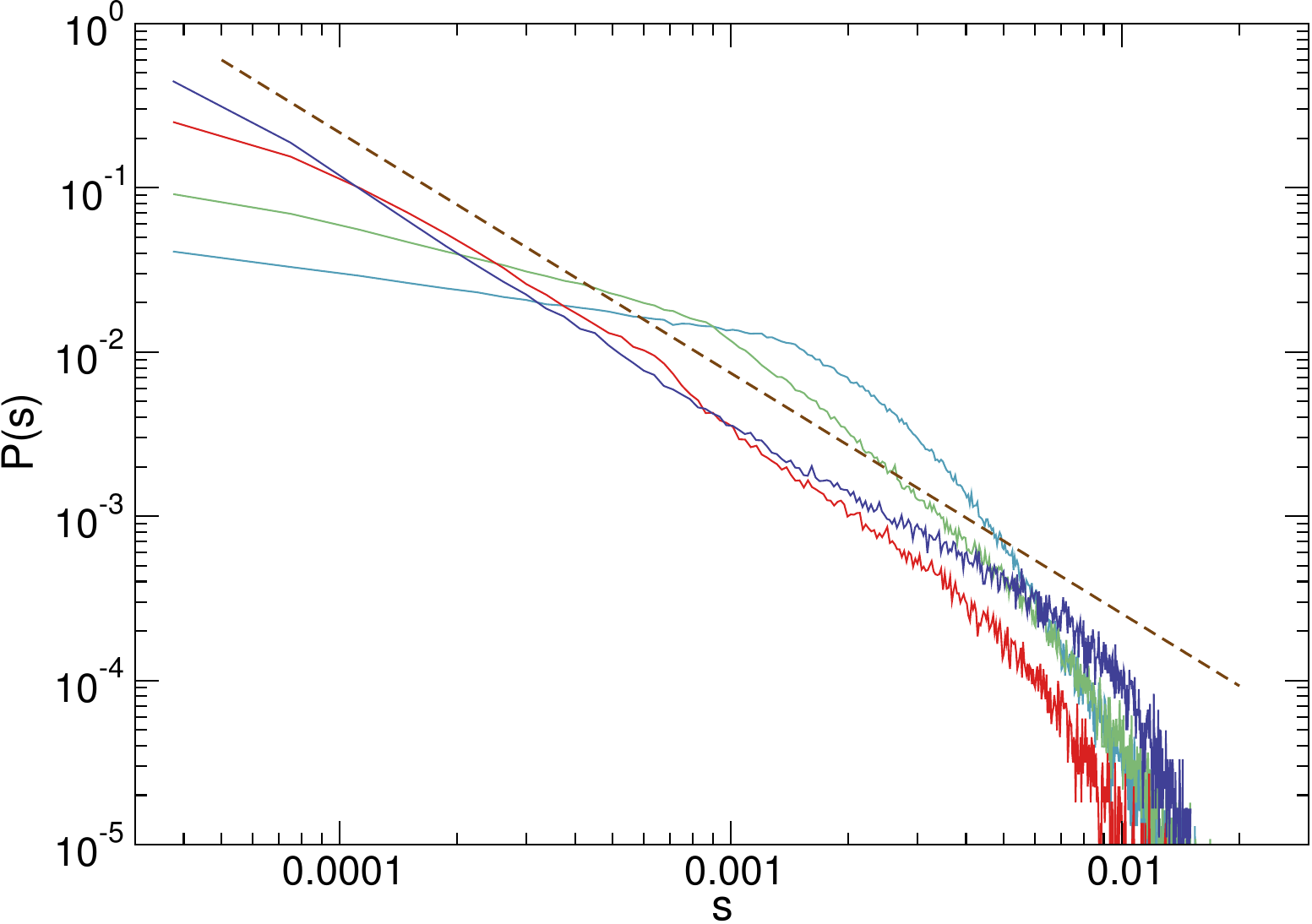}
\caption{
  $P(s)$
  for a system with $\phi_{\rm obs} = 0.1727$ and $l_{r} = 320$
  at $\phi_{\rm tot} = 0.377$ (light blue), $0.5$ (green), $0.628$ (red), and
  $0.754$ (dark blue). 
The dashed line is a power law fit with exponent $\beta = 1.465$.
}
\label{fig:7}
\end{figure}

In Fig.~\ref{fig:7} we show $P(s)$
for a system with fixed $\phi_{\rm obs} = 0.1727$ and $l_{r} = 320$ 
at $\phi_{\rm tot} = 0.377$, $0.5$, $0.628$, and $0.754$, along with a power law
fit with exponent $\beta=1.465$.
For $\phi_{\rm tot} = 0.377$ and $\phi_{\rm tot}=0.5$,
$P(s)$ cannot be fit by a single power law,
while for $\phi_{\rm tot} = 0.65$ and $\phi_{\rm tot}=0.754$, there
is good agreement with the power law fit,
indicating that critical behavior only appears when the system is not
too sparse.

\section{Discussion}
We ask where the criticality 
we observe in our active matter system originates.
When $l_{r}$ and $\phi_{a}$ are large enough,
the obstacle-free system enters a phase separated state in which
the dense cluster regions have a density close to the jamming
density, so the dense regions can be viewed as an assembly of
grains that is close to the critical Point J identified
in Ref.~\cite{16}.
Several studies of yielding in two-dimensional foams \cite{35} 
and granular matter \cite{36}
identify avalanches that have a power law size distribution with an
exponent of 
$\beta = 1.5$,
while other simulations of yielding in two-dimensional granular matter
give avalanches with power law size distribution exponents of
$\beta=1.43$ \cite{37}.
The exponent we observe
is close to the
value $\beta=1.5$ 
predicted using mean field models \cite{38,39}.
There are also studies of yielding in soft particulate matter systems
in which avalanches with exponents
$\beta=1.35$  appear \cite{40},
while experiments on frictional granular matter give avalanche
exponents of
$\beta= 1.24$ \cite{37}. 
We argue that in the thermal limit of small $l_r$, our
active disks behave like a liquid with short correlation lengths,
so portions of the system can readily flow as long as there is space for
motion between the obstacles.
When $l_r=0$ in the limit of zero activity,
the disks reach
a completely clogged state where no fluctuations and therefore no
avalanches occur.
At large $l_r$, the disks self-cluster and locally behave 
like a granular solid just on the verge of jamming, where large correlation lengths
emerge, but
the fact that the disks are active and are always attempting to move
prevents the system from become permanently trapped in 
a jammed state.
Instead, occasional activity-induced unjamming events occur
that have the appearance of avalanches.
The motion of our active disks
is impeded by the presence of obstacles; however,
even in the absence of 
obstacles, the active clusters can undergo local rearrangements
that can occur suddenly as an avalanche.
In previous simulations
of active disks without obstacles,
local velocity
fluctuations of a single driven probe disk were power law distributed
with an exponent of $\beta = 2.0$ when the activity
was large enough to permit self-clustering
to occur, while in the low activity limit
where the system acts like a uniform fluid,
the velocity fluctuation distribution was exponential \cite{34}.

In this work we focus on the case where there is an applied drive in order
to characterize the avalanche behavior;
however, it would also be interesting to study
a dense active disk assembly in the presence of quenched disorder rather
than obstacles to see whether avalanches also occur
in the absence of
an external drift force.
Previous numerical simulations of active disks on quenched
disorder have only explored the low density regime \cite{41}.
It would also be interesting to determine whether other active matter models
such as flocking particles
exhibit avalanche behavior in the presence of quenched disorder.
Simulations have already shown nonmonotonic transport behavior
in such systems \cite{42} as well as disorder-induced
transitions from flocking to non-flocking states \cite{43},
and there are now experimental realizations of
flocking systems with quenched disorder \cite{44} that could be used to
explore this issue.

\section{Summary}
We have numerically examined the avalanche
behavior of active matter
composed of run-and-tumble disks driven through 
a random obstacle array.
We measure avalanche sizes in terms of the average instantaneous
velocity of the active disks.
At low activity the system becomes trapped in a completely clogged state, while 
at intermediate levels of activity the
disks act like a fluid that can flow continuously
among the obstacles, producing a bimodal avalanche size distribution.
At large run lengths, the disks undergo
self-clustering and their motion becomes
highly intermittent,
taking the form of avalanches of correlated disk motion that have
a power law size distribution  with an exponent of
$\beta = 1.465$. 
We argue that the intermittency
results from self-clustering, which causes the system to
act like a granular solid that is near the jamming point, 
and that the activity-induced avalanches are similar to the behavior
observed in the yielding of marginally stable solids
such as foams or granular packings,
where avalanches with similar size distribution power law exponents appear.
Finally, we find that when the density
of obstacles is large enough,
the avalanche size distribution is cut off at large sizes, suggesting that
there is a critical disorder density that maximizes the critical nature of the
avalanches.
Our results indicate that activity provides
another route for creating critical nonequilibrium states in particulate matter.

\ack
This work was carried out under the auspices of the 
NNSA of the U.S. DoE at LANL under Contract No.
DE-AC52-06NA25396.


\section*{References}
\begin{thebibliography}{9}

\bibitem{1}
  Olson C J, Reichhardt C and Nori F 1997
  Superconducting vortex avalanches, voltage bursts, and vortex plastic flow:
  Effect of the microscopic pinning landscape on the macroscopic properties
{\it Phys. Rev. B} {\bf 56} 6175

\bibitem{2}
  Bassler K and Paczuski M 1998
Simple model of superconducting vortex avalanches
{\it Phys. Rev. Lett.} {\bf 81} 3761

\bibitem{3}
  Altshuler E 2004
Experiments in vortex avalanches
{\it Rev. Mod. Phys.} {\bf 76} 471

\bibitem{4}
  Zapperi S, Cizeau P, Durin G and Stanley H E 1998
  Dynamics of a ferromagnetic domain wall: Avalanches, depinning transition, and the
  Barkhausen effect
  {\it Phys. Rev. B} {\bf 58} 6353

\bibitem{5}
  Sethna J P, Dahmen K and Myers C R 2001
  Crackling noise
  {\it Nature} {\bf 410} 242

\bibitem{6}
  Carlson J M and Langer J S 1989
Properties of earthquakes generated by fault dynamics
{\it Phys. Rev. Lett.} {\bf 62} 2632

\bibitem{7}
  Fisher D S 1998
  Collective transport in random media: From superconductors to earthquakes
  {\it Phys. Rep.} {\bf 301} 113

\bibitem{8}
  Reichhardt C and Reichhardt C J O 2017
  Depinning and nonequilibrium dynamic phases of particle assemblies driven over
  random and ordered substrates: A review
  {\it Rep. Prog. Phys.} {\bf 80} 026501

\bibitem{9}
  Miguel M-C, Vespignani A, Zapperi S, Weiss J and Grasso J-R 2001
  Intermittent dislocation flow in viscoplastic deformation
  {\it Nature} {\bf 410} 667

\bibitem{10}
  Zaiser M 2006
Scale invariance in plastic flow of crystalline solids
{\it Adv. Phys.} {\bf 55} 185 

\bibitem{11}
McDermott D, Reichhardt C J O and Reichhardt C 2016
Avalanches, plasticity, and ordering in colloidal crystals under compression
{\it Phys. Rev. E} {\bf 93} 062607

\bibitem{12}
  Salerno K M, Maloney C E and Robbins M O 2012
Avalanches in strained amorphous solids: Does inertia destroy critical behavior?
{\it Phys. Rev. Lett.} {\bf 109} 105703 

\bibitem{13}
  Lin Jie, Gueudre T, Rosso A and Wyart M 2015
Criticality in the approach to failure in amorphous solids
{\it Phys. Rev. Lett.} {\bf 115} 168001

\bibitem{14}
  Regev I, Weber J, Reichhardt C, Dahmen K A and Lookman T 2015
  Reversibility and criticality in amorphous solids
  {\it Nature Commun.} {\bf 6} 8805

\bibitem{15}
  Nicolas A, Ferrero E E, Martens K and Barrat J-L 2017
    Deformation and flow of amorphous solids: a review of mesoscale elastoplastic models
  {\it Preprint} arXiv:1708.09194

\bibitem{16}
  Liu A J and Nagel S R 1998
  Nonlinear dynamics: Jamming is not just cool any more
  {\it Nature} {\bf 396} 21

\bibitem{17}
  Reichhardt C and Reichhardt C J O 2014
Aspects of jamming in two-dimensional athermal frictionless systems
{\it Soft Matter} {\bf 10} 2932

\bibitem{18}
  Drocco J A, Hastings M B, Reichhardt C J O and Reichhardt C 2005
Multiscaling at point J: Jamming is a critical phenomenon
{\it Phys. Rev. Lett.} {\bf 95} 088001

\bibitem{19}
  Candelier R and Dauchot O 2010
  Journey of an intruder through the fluidization and jamming transitions of a
  dense granular media
{\it Phys. Rev. E} {\bf 81} 011304

\bibitem{20}
  Reichhardt C J O and Reichhardt C 2010
  Fluctuations, jamming, and yielding for a driven probe particle in disordered
  disk assemblies
{\it Phys. Rev. E} {\bf 82} 051306

\bibitem{21}
  Ar{\' e}valo R and Ciamarra M P 2014
Size and density avalanche scaling near jamming
{\it Soft Matter} {\bf 10} 2728

\bibitem{22}
  Woldhuis E, Chikkadi V, van Deen M S, Schall P and van Hecke M 2015
Fluctuations in flows near jamming
{\it Soft Matter} {\bf 11} 7024

\bibitem{23}
  Franz S and Spigler S 2017
Mean-field avalanches in jammed spheres
{\it Phys. Rev. E} {\bf 95} 022139

\bibitem{24}
  Ramaswamy S 2010
  The mechanics and statistics of active matter
  {\it Annu. Rev. Condens. Matter Phys.} {\bf 1} 323

\bibitem{25}
  Bechinger C, Di Leonardo R, L{\" o}wen H, Reichhardt C, Volpe G and Volpe G 2016
  Active Brownian particles in complex and crowded environments
  {\it Rev. Mod. Phys.} {\bf 88} 045006

\bibitem{26}
  Fily Y and Marchetti M C 2012
  Athermal phase separation of self-propelled particles with no alignment
  {\it Phys. Rev. Lett.} {\bf 108} 235702

\bibitem{27}
  Redner G S, Hagan M F and Baskaran A 2013
  Structure and dynamics of a phase-separating active colloidal fluid
  {\it Phys. Rev. Lett.} {\bf 110} 055701

\bibitem{28}
  Cates M A and Tailleur J 2015
  Motility-induced phase separation
  {\it Annu. Rev. Condens. Mat. Phys.} {\bf 6} 219

\bibitem{29}
  Palacci P, Sacanna S, Steinberg A P, Pine D J, and Chaikin P M 2013
  Living crystals of light-activated colloidal surfers
  {\it Science } {\bf 339} 936

\bibitem{30}
  Buttinoni I, Bialk{\' e} J, K{\" u}mmel F, L{\" o}wen H, Bechinger C and Speck T 2013
  Dynamical clustering and phase separation in suspensions of self-propelled
  colloidal particles
{\it Phys. Rev. Lett.} {\bf 110} 238301

\bibitem{31}
  Reichhardt C and Reichhardt C J O 2014
Active matter transport and jamming on disordered landscapes
{\it Phys. Rev. E} {\bf 90} 012701

\bibitem{32}
  Perkovic O, Dahmen K and Sethna J P 1995
  Avalanches, Barkhausen noise, and plain old criticality
  {\it Phys. Rev. Lett.} {\bf 75} 4528

\bibitem{33}
  Chern G-W, Reichhardt C and Reichhardt C J O 2014
Avalanches and disorder-induced criticality in artificial spin ices
{\it New J. Phys} {\bf 16} 063041

\bibitem{34}
Reichhardt C and Reichhardt C J O 2015
Active microrheology in active matter systems: Mobility, intermittency, and avalanches
{\it Phys. Rev. E} {\bf 91} 032313

\bibitem{hermann}
  Luding S and Herrmann H J 1999
  Cluster growth in freely cooling granular media
  {\it Chaos} {\bf 9} 673

\bibitem{35}
  Kabla A, Scheibert J and Debregeas G 2007
Quasistatic rheology of foams: II. Transition to shear banding
{\it J. Fluid Mech.} {\bf 587} 45

\bibitem{36}
  Hayman N W, Ducloue L, Foco K L and Daniels K E 2011
  Granular controls on periodicity of stick-slip events: kinematics and
  force-chains in an experimental fault
  {\it Pure Appl. Geophys.}  {\bf 168} 2239

\bibitem{37}
  Bar{\' e}s J, Wang D, Wang D, Bertrand T, O'Hern C S and Behringer R P 2017
    Local and global avalanches in a 2D sheared granular medium
  {\it Preprint}  arXiv:1709.01012 

\bibitem{38}
  Dahmen K,  Ben-Zion Y and Uhl J T 2009
  Micromechanical model for deformation in solids with universal predictions for
  stress strain curves and slip avalances
  {\it Phys. Rev. Lett.} {\bf 102} 175501

\bibitem{39}
  Salje E K H and Dahmen K A 2014
  Crackling noise in disordered materials
  {\it Ann. Rev. Condens. Mat. Phys.} {\bf 5} 233

\bibitem{40}
  Lin J, Lerner E, Rosso A and Wyart M 2014
  Scaling description of the yielding transition in soft amorphous solids at
  zero temperature
  {\it Proc. Natl. Acad. Sci. (USA)}  {\bf 111} 14382 

\bibitem{41}
  Zeitz M, Wolff K and Stark H 2017
Active Brownian particles moving in a random Lorentz gas
{\it Eur. Phys. J.E} {\bf 40} 23

\bibitem{42}
  Chepizhko O, Altmann E G and Peruani F 2013
  Optimal noise maximizes collective motion in heterogeneous media
  {\it Phys. Rev. Lett.} {\bf 110} 238101

\bibitem{43}
  Quint D and Gopinathan A 2015
  Topologically induced swarming phase transition on a 2D percolated lattice
  {\it Phys. Biol.} {\bf 12} 046008

\bibitem{44}
  Morin A, Desreumaux N, Caussin J-B and Bartolo D 2017
  Distortion and destruction of colloidal flocks in disordered environments
  {\it Nature Phys.} {\bf 13} 63

\end{thebibliography}
\end{document}